\documentclass[prl,twocolumn,showpacs,amsmath,amssymb]
{revtex4}

\usepackage[dvips]{color}
\usepackage{graphicx}

\begin{document}

\title{Fermionization in an expanding 1D gas of hard-core bosons}

\author{Marcos Rigol}
\affiliation{Institut f\"ur Theoretische Physik III, Universit\"at 
Stuttgart, Pfaffenwaldring 57, D-70550 Stuttgart, Germany.}
\author{Alejandro Muramatsu}
\affiliation{Institut f\"ur Theoretische Physik III, Universit\"at 
Stuttgart, Pfaffenwaldring 57, D-70550 Stuttgart, Germany.}

\begin{abstract}
We show by means of an exact numerical approach that the momentum 
distribution of a free expanding gas of hard-core bosons on a 
one-dimensional lattice approaches to the one of noninteracting 
fermions, acquiring a Fermi edge. Yet there is a power-law decay of 
the one-particle density matrix $\rho_x\sim 1/\sqrt{x}$, as usual 
for hard-core bosons in the ground state, which accounts for a large 
occupation of the lowest natural orbitals for all expansion times. 
The fermionization of the momentum distribution function, 
which is not observed in equilibrium, is analyzed in detail. 

\end{abstract}

\pacs{03.75.Kk, 03.75.Lm, 05.30.Jp}
\maketitle

At very low temperatures and densities a one-dimensional (1D) gas of bosons 
is expected to behave as a gas of impenetrable particles known as hard-core 
bosons (HCB) \cite{olshanii98}. Two recent experiments successfully  
achieved the required parameter regime and made HCB a physical 
reality \cite{paredes04,kinoshita04}. In contrast to bosons in higher 
dimensions, 1D HCB share many properties with noninteracting spinless 
fermions to which they can be mapped \cite{girardeau60}. Thermodynamic 
properties like the total energy, and microscopic properties like density 
profiles are identical in both systems. On the contrary, quantities like 
the momentum distribution function ($n_k$) \cite{lenard64} and the so-called 
natural orbitals (NO) \cite{penrose56} are very different for HCB and spinless 
fermions. This is due to the different behavior of the off-diagonal 
elements of the one-particle density matrix (OPDM) in both systems 
(see {\it e.g.} Ref.\ \cite{cazalilla04}). 

An important point for the comparison between experimental results and 
theory in Refs.\ \cite{paredes04,kinoshita04} has been the awareness of 
the effects of the trapping potential in the properties of the HCB gas.
On a harmonic trap it has been found that the power-law decay of the 
OPDM $\rho_{ij}\sim |x_i-x_j|^{-1/2}$, known from homogeneous systems 
\cite{lenard64}, is renormalized by a factor that depends on the density 
at points $i$ and $j$. This factor is proportional to $[n_in_j]^{1/4}$ 
for continuous systems \cite{forrester03}, and to 
$[n_i(1-n_i)n_j(1-n_j)]^{1/4}$ for HCB on a lattice \cite{rigol04_1}. 
Furthermore, the power-law decay of the OPDM has been found to be universal 
independently of the power of the confining potential \cite{rigol04_1}. 
Even for systems out of equilibrium, that start their evolution from a totally 
uncorrelated state, the power law above develops dynamically and leads 
to the emergence of quasi-condensates at finite momentum \cite{rigol04_2}.

In this work we show that during the free expansion of 
1D HCB in a lattice a further degree of fermionization takes place, 
i.e., $n_k$ of the HCB becomes for long expansion times equal to the one 
of noninteracting fermions, displaying a Fermi edge.
This feature, absent in equilibrium, can be easily confirmed 
experimentally with a setup like the one in Ref.\ \cite{kinoshita04} 
where the expansion of the 1D gas was studied after removing the axial 
confinement. Other quantities like the OPDM and the NO still evidence the 
strongly interacting character of the HCB system.
  
We obtain the exact dynamics of expanding clouds of HCB on 1D 
lattices on the basis of the Jordan-Wigner transformation, which
maps the HCB Hamiltonian into the one of spinless fermions \cite{rigol04_2} .
Apart from being exact, our method allows to consider relatively large number 
of particles and system sizes, which can be even larger than the ones 
in the present experimental setups. Results for continuous systems can be 
extrapolated from very low densities in the lattice \cite{rigol04_1}.
At time $\tau=0$ we switch off the trapping potential, and 
start the time evolution from the ground state of the HCB Hamiltonian
\begin{equation}
\label{HamHCB} H_{HCB} = -t \sum_{i} ( b^\dagger_{i} b^{}_{i+1}
+ \text{H.c.} ) + V_2 \sum_{i} x_i^2\ n_{i },
\end{equation}
which has the additional on-site constraints $b^{\dagger 2}_{i}= b^2_{i}=0$, 
$\left\lbrace  b^{}_{i},b^{\dagger}_{i}\right\rbrace =1$. 
In Eq.\ (\ref{HamHCB}), $b^{\dagger}_{i}$ and $b_{i}$ represent the 
HCB creation and annihilation operators, respectively, 
$n_{i }= b^{\dagger}_{i}b^{}_{i}$ the particle number operator, 
$t$ the hopping parameter, and $V_2$ the curvature of the 
harmonic confining potential. 

In order to characterize the initial state, we use the characteristic 
density $\tilde{\rho}=N_ba/\zeta$ \cite{rigol04_1,rigol04_2},
which for trapped systems plays a similar role than the density in 
periodic systems. $N_b$ is the number of HCB,  
$\zeta=\left( V_2/t\right)^{-1/2}$ is a length scale of the trap in the 
presence of the lattice, and $a$ is the lattice constant. 
The quasi-momentum distribution function $n_k$ is also normalized 
by the length scale $\zeta$ as 
$n_k=(a/\zeta)\sum_{ij} e^{-ik(i-j)}\rho_{ij}$, with 
$\rho_{ij}=\langle b^\dagger_{i}b_{j}\rangle$.
In addition, we define the Fermi momentum $k_F$ associated to the fermions 
as $\epsilon_F=-2t\cos(k_Fa)$, where $\epsilon_F$ is the energy of the last 
occupied fermionic single-particle state in the trap at $\tau=0$. 
(Although in the trap $n_k$ is continuous at $k_F$, it is 
possible to see that $n_k$ approaches zero even faster than exponentially 
for $k>k_F$.)

\begin{figure}[h]
\includegraphics[width=0.36\textwidth,height=0.25\textwidth]
{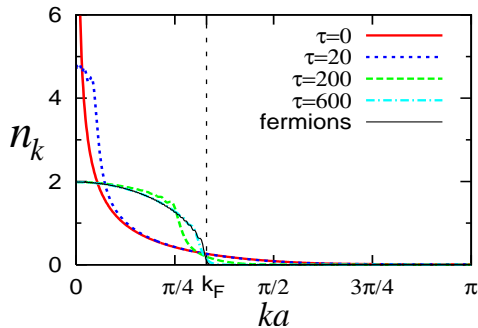} \vspace{-0.2cm}
\caption{(color online). $n_k$ for 100 HCB expanding from 
an initial state with $\tilde{\rho}=0.51$, compared to the one 
for the corresponding fermions. Times ($\tau$) are given in units 
of $\hbar/t$, and $k_F$ denotes de Fermi momentum as defined in the text.}
\label{PerfilK}
\end{figure}

In Fig.\ \ref{PerfilK} we show $n_k$ for an expanding gas of HCB 
at four different times, and compare it with the one of noninteracting 
fermions (which does not change during the expansion). 
Several issues are evident:
i) Shortly after switching off the trapping potential, the peak at 
$n^b_{k=0}$ disappears. 
ii) For $k<k_F$ a redistribution of the population of $k$ states takes
place, such that starting from $k \sim 0$, $n_k$ of HCB matches 
in time the one for fermions.
iii) $k$ states with $k>k_F$ become 
less populated and an edge develops at $k_F$. The overall process leads 
to an $n_k$ for the HCB that is equal to the one of the fermions.

\begin{figure}[h]
\includegraphics[width=0.38\textwidth,height=0.25\textwidth]
{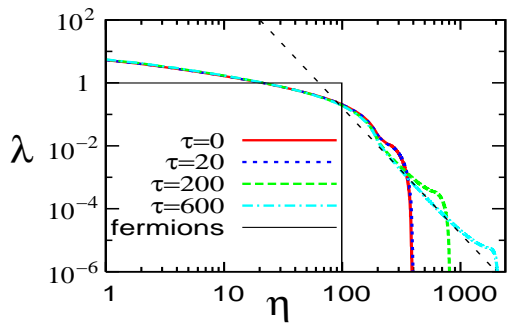} \vspace{-0.3cm}
\caption{(color online). Occupation of the NO vs $\eta$ for the same 
initial trap parameters and times of Fig.\ \ref{PerfilK}. The thin dashed 
line corresponds to a power law $\eta^{-4}$, which is known from 
equilibrium systems at low densities \cite{rigol04_1}.}
\label{NO}
\end{figure}

After observing the fermionization of $n_k$ for HCB, one could naively expect 
that something similar could be happening with the NO occupations. 
The NO ($\phi^\eta_i$) are the eigenfunctions of the OPDM 
$\sum_j \rho_{ij}\phi^\eta_j= \lambda_{\eta}\phi^\eta_i$ \cite{penrose56}, 
i.e., they are effective single-particle states with occupations 
$\lambda_{\eta}$. For noninteracting fermions the NO are the 
eigenfunctions of the Hamiltonian and their occupation is one. 
We find that in contrast to $n_k$ the NO occupations do not 
fermionize, as can be seen in Fig.\ \ref{NO} for the same 
parameters of Fig.\ \ref{PerfilK}.

\begin{figure}[h]
\includegraphics[width=0.37\textwidth,height=0.245\textwidth]
{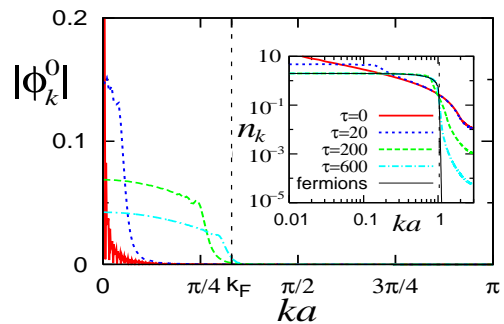} \vspace{-0.26cm}
\caption{(color online). Fourier transform of the lowest NO, and 
$n_k$ in a logarithmic scale (inset), for the same 
initial trap parameters and times of Figs.\ \ref{PerfilK} and 
\ref{NO}. $k_F$ (vertical dashed line) denotes de Fermi momentum.}
\label{NOK}
\end{figure}

There are two features of the NO occupations that we find worth noticing. 
The first one, which is difficult to distinguish in Fig.\ \ref{NO}, is that 
the lowest natural orbitals slightly increase their occupations during the 
expansion of the gas. This can be intuitively understood as an increase 
of the ``coherence'' of the system due to an increase of the system size, 
which delocalize the HCB over more lattice sites. Something similar occurs 
in the ground state occupations of the lowest natural orbitals when for the 
same number of particles the curvature of the trap is decreased 
\cite{rigol04_1}. However, in equilibrium systems, $n_{k=0}$ also 
increases with the increase of $\lambda_0$. The apparent contradiction 
between the decrease of $n_{k=0}$ and the increase of $\lambda_0$ in 
the expanding gas can be resolved observing the Fourier transform of the 
lowest NO at $\tau=0$ and $\tau>0$. As it can be seen in Fig.\ \ref{NOK}, 
initially $|\phi^0_k|$ has a peak at $k=0$ showing that quasi-condensation 
occurs around $k=0$, and this is reflected in $n_k$. For $\tau>0$ the 
lowest NO becomes an extended object in $k$-space so that the HCB forming 
the quasi-condensate have many different momenta, basically as many as $n_k$ 
in Fig.\ \ref{PerfilK}. Hence, there is no contradiction between the observed 
behavior of the NO occupations and $n_k$, although the last one is clearly 
different to the one of systems in equilibrium.

The second feature that is worth noticing, sets in only when the density 
of the expanding HCB becomes very low. Then the universal power-law decay 
of the NO occupations for large-$\eta$ ($\lambda_\eta\sim A_{N_b}\eta^{-4}$), 
known from equilibrium systems \cite{rigol04_1}, also appears in 
nonequilibrium (Fig.\ \ref{NO}). In Ref.\ \cite{rigol04_1}, the prefactor 
of the power-law $A_{N_b}$ was found to depend only on $N_b$ independently 
of the confining potential. We find out of equilibrium that $A_{N_b}$ is
exactly the same than in the ground state case \cite{rigol04_1}. 
For $N_b=100$ we have plotted $\lambda_\eta= A_{N_b}\eta^{-4}$ in 
Fig.\ \ref{NO}. The long tail of the momentum distribution function 
$n_k\sim |k|^{-4}$, that in equilibrium appears together with the 
$\lambda_\eta\sim \eta^{-4}$ \cite{rigol04_1}, is in general 
not present in nonequilibrium since for large $\tau$, 
$n_k$ for HCB starts to behave like the one of fermions 
(inset in Fig.\ \ref{NOK}).

Considering the previous results for the behavior of the NO 
occupations, which is similar to the one known in equilibrium
\cite{rigol04_1}, one expects that also the OPDM should behave similarly. 
Since in nonequilibrium $\rho_{ij}=|\rho_{ij}|e^{i\theta_{ij}}$ 
is in general a complex object, in order to compare with equilibrium 
systems we first study its modulus. Results for the same systems of 
Figs.\ \ref{PerfilK}--\ref{NOK} are shown in Fig.\ \ref{DensMatrix}(a). 
Figure \ref{DensMatrix}(a) shows that $|\rho_{ij}(\tau)|$ have exactly the 
same form than $\rho_{ij}$ in equilibrium systems \cite{rigol04_1}. 
For large values of $|x_i-x_j|$ a power-law decay 
$|\rho_{ij}|\sim|x_i-x_j|^{-1/2}$ can be observed {\it for all times}, 
and the prefactor of the power law decreases with the reduction of the 
local densities in the system. Hence, the slow decay of the 
one-particle correlations is the one accounting for the large 
($\sim\sqrt{N_b}$) occupation of the lowest NO. 
On the other hand, Figs.\ \ref{DensMatrix}(b)-(d) show that the phase of 
the OPDM ($\theta_{ij}$) starts to increasingly oscillate at large 
distances. 
In particular Fig.\ \ref{DensMatrix}(b) shows that after 
a very short time, when the modulus of the OPDM have almost not 
changed, $\theta_{ij}$ have started to oscillate for $|x_i-x_j|\gg a$ 
producing a fast destruction of the zero momentum peak in $n^b_k$, as shown 
in Fig.\ \ref{PerfilK}.

\begin{widetext}

\vspace{-0.8cm}

\begin{figure}[b]
\includegraphics[width=0.66\textwidth,height=0.26\textwidth]
{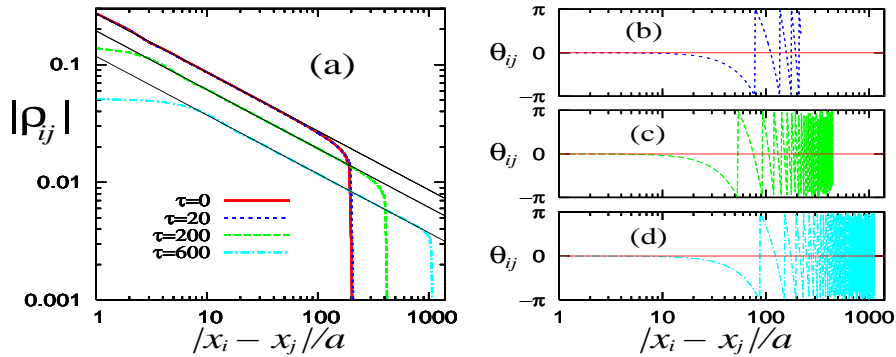} \vspace{-0.3cm}
\caption{(color online). Modulus of the OPDM (a) and its phase (b)-(d) 
for the same initial trap parameters and times of Figs.\ 
\ref{PerfilK}--\ref{NOK}. Both quantities have been evaluated with 
respect to the middle of the expanding cloud of HCB.  
Thin continuous lines in (a) correspond to power laws $|x_i-x_j|^{-1/2}$.}
\label{DensMatrix}
\end{figure}

\end{widetext}

In order to gain further insight into the fermionization described above, 
we observe that for noninteracting particles, 
given the initial one-particle density matrix $\rho_{ab}$, the density 
profile at time $\tau$ can be evaluated as \cite{pedri04_1}  
\begin{equation}
n_i(\tau)=\sum_{ab}G^{*}_{i,a}(\tau)G_{i,b}(\tau) \rho_{ab},
\label{density1}
\end{equation}
where the free one-particle propagator can be written in the form 
$G_{i,j}(\tau)=\sum_k e^{-i\tau/\hbar[\epsilon_k-\hbar k(x_i-x_j)/\tau]}$, 
$\epsilon_k$ being the dispersion relation. Fourier transforming the 
initial OPDM into $\varrho_{k_1,k_2}$ and performing straightforward 
integrations, Eq.\ (\ref{density1}) can be rewritten as
\begin{eqnarray}
n_i(\tau)&=&\sum_{k_1,k_2} \varrho_{k_1k_2} 
e^{i\tau/\hbar[\epsilon_{k_1}-\hbar k_1x/\tau]}
e^{-i\tau/\hbar[\epsilon_{k_2}-\hbar k_2x/\tau]} \nonumber \\
&\simeq& \dfrac{1}{\tau \epsilon''_{k=k_0(x_i/\tau)}}
\varrho_{k_1=k_0(x_i/\tau),k_2=k_0(x_i/\tau)},
\label{perfil2}
\end{eqnarray} 
where in the last step we have assumed $\tau$ to be very large and 
performed the summations using the saddle point approximation. 
[$k_0(x_i/\tau)$ is determined by the expression 
$\epsilon'_{k_0}=\hbar x_i/\tau$, where the prima means $k$-derivative.]
Equation (\ref{perfil2}) shows that for $\tau\rightarrow \infty$ the density 
profile is only determined by a rescaling of the diagonal part of 
$\varrho_{k_1,k_2}$, which is nothing but $n_k$. In the continuum limit 
$\epsilon_k=\hbar^2k^2/2m$ and $n(x,\tau)\simeq (m/\tau) 
\tilde{n}(k=m x/\hbar \tau)$ [$\tilde{n}(k)$ is the momentum distribution 
function], which is a known result that can be also obtained by other means 
\cite{pedri04_1}. Since the arguments above are valid for both 
fermions and bosons, the fact that $n_k$ of HCB converges to that 
of fermions after expansion could be explained at this point 
{\em if after a certain time, the 
expanding HCB could be considered as non-interacting}. 
This would mean that $n_k$ for HCB would be determined by a rescaling 
of the density profile that, on the other hand, both in the continuum 
\cite{santos02} and on the lattice \cite{rigol04_1,rigol04_2}
is the same as that of fermions due to the mapping connecting both.  

However, a non-interacting treatment of the HCB expansion is invalidated 
by Fig.\ 4(a). There it is shown that at all times, even after 
fermionization, the density matrix decays as $1/\sqrt{x}$, a power 
that corresponds to HCB, and hence, the system corresponds to strongly 
interacting particles. Therefore, the expansion out of equilibrium 
leads to a new kind of bosonic state, with a Fermi edge  
in the momentum distribution function but still the effective 
one-particle states, as given by the natural orbitals, exhibit a 
high occupation as expected for bosons.
 
\begin{figure}[h]
\includegraphics[width=0.48\textwidth,height=0.22\textwidth]
{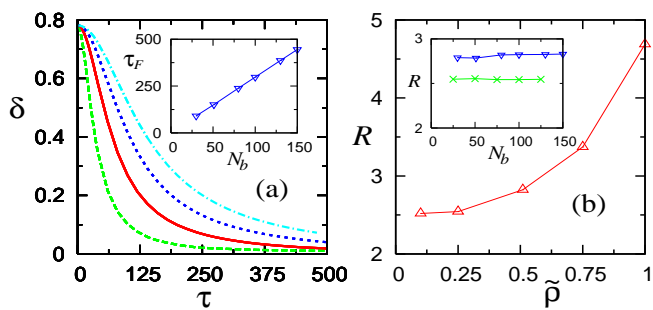} \vspace{-0.6cm}
\caption{(color online). Fermionization of $n^b_k$ during the expansion
of the gas. (a) Decrease of $\delta$ (see text) as a function of time 
for $\tilde{\rho}=0.51$; $N_b=51$ (dashed line), $N_b=100$ (continuous line), 
$N_b=150$ (dotted line), and $N_b=200$ (dashed-dotted line).
The inset shows the fermionization time $\tau_F$ (see text) vs $N_b$, 
for $\tilde{\rho}=0.51$.
(b) Ratio $R$ between the size of gas for $\delta=0.05$ and its 
original size vs  $\tilde{\rho}$, $N_b=100$. The inset shows 
$R$ vs $N_b$ for $\tilde{\rho}=0.25$ (\textcolor{green}{$\times$}) 
and $\tilde{\rho}=0.51$ (\textcolor{blue}{$\nabla$}).}
\label{Fermionization}
\end{figure}

In what follows we study how exactly the fermionization of the 
momentum distribution for bosons $n^b_k$ occurs in time, and how it 
depends on experimental parameters like number of particles and 
characteristic densities. For that we analyze the relative area 
between $n^b_k$ and the momentum distribution for fermions
$n^f_k$, $\delta=(\sum_k |n^b_k-n^f_k|)/(\sum_k n^b_k)$. This is shown 
in Fig.\ \ref{Fermionization}(a) for $\tilde{\rho}=0.51$ and different 
fillings of the trap. Fig.\ \ref{Fermionization}(a) shows that changes 
in $n^b_k$ occur fast in terms of the characteristic time of the system, 
which is given by $\hbar/t$. In addition, if one chooses a criterion
like $\delta=0.05$ to state that $n_k^b$ has fermionized, it is possible to 
see in the inset of Fig.\ \ref{Fermionization}(a) that for a given 
characteristic density the fermionization time ($\tau_F$) grows linearly 
with the number of HCB in the trap.

Another question that is important to answer is the 
consequence of changing $\tilde{\rho}$ in the ground state of the trap.
In order to compare systems with different $\tilde{\rho}$, i.e., 
different $n_k$, we display in Fig.\ \ref{Fermionization}(b) the ratio $R$ 
between the size of the cloud once $\delta=0.05$ and its initial size. 
Fig.\ \ref{Fermionization}(b) shows that with decreasing $\tilde{\rho}$ 
the ratio $R$ reduces up to $\sim 2.5$, and that for $\tilde{\rho}>0.5$ 
it increases very fast. For low $\tilde{\rho}$, such that the 
averaged interparticle distance is much larger than the lattice spacing, 
the initial lattice gas is equivalent to the one in continuous 
systems. This means that a fermionized $n^b_k$ will be more easily 
observed in continuous systems \cite{kinoshita04} than in the lattice
\cite{paredes04}. (In the continuous case, the asymptotic fermionization 
of $n^b_k$ was obtained previously in Ref.\ \cite{sutherland98}.)
In addition, the inset in Fig.\ \ref{Fermionization}(b) 
shows that the ratio $R$ remains basically constant for a given 
characteristic density when the number of particles in the trap in 
changed. 

\begin{figure}[t]
\includegraphics[width=0.36\textwidth,height=0.24\textwidth]
{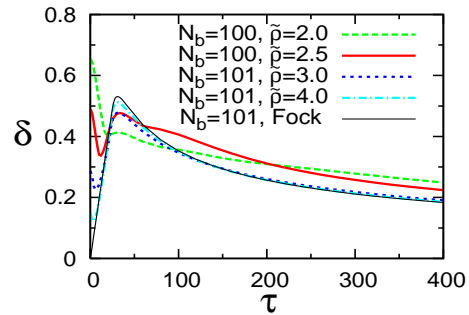} \vspace{-0.25cm}
\caption{(color online). Fermionization of $n^b_k$ during the expansion
of the gas. For $\tilde{\rho}=3.0$ and 4.0 a Mott insulator is formed 
in the middle of the trap. The behavior for a pure Fock state (Fock) 
with one particle per lattice site is also shown.}
\label{FermionizationMI}
\end{figure}

Increasing the characteristic density of the initial system 
beyond the values in Fig.\ \ref{Fermionization}(b) one starts observing 
a behavior of $\delta$ which is different to the one seen in Fig.\ 
\ref{Fermionization}(a). The reason is that particles become more 
localized in the middle of the trap, and eventually after 
$\tilde{\rho}\sim 2.6-2.7$ a Mott insulator appears in the system.
This localization effect also generates an $n^b_k$ which approaches to the 
one of the fermions in the initial state. [In the limit of all lattice sites 
with occupation one \cite{rigol04_2}, $n^b_k(\tau=0)=n^f_k$.] When 
such systems are released from the trap, quasi-long range correlations 
start to develop between initially uncorrelated particles and they lead 
to the formation of traveling quasi-condensates \cite{rigol04_2}. This 
generates an $n^b_k(\tau)$ at short times that may be more different to 
$n^f_k$ than $n^b_k(\tau=0)$, as it can be seen in Fig.\ 
\ref{FermionizationMI}. However, after long times one can see that a 
fermionization of $n^b_k$ starts to occurs as before for smaller 
$\tilde{\rho}$. One should notice that as shown in Fig.\ 
\ref{FermionizationMI}, the time scales for the fermionization 
process for large $\tilde{\rho}$ are very long, and always start 
affecting the low-momenta region first, so that still the dynamically 
generated quasi-condensates, which have $k\sim \pm \pi/2$, 
could be used as atom lasers \cite{rigol04_2}.

In summary, we have shown that during the expansion of a 1D gas of HCB 
the momentum distribution function becomes equal to the one of the 
equivalent noninteracting fermions. This is an effect that can be seen 
experimentally in systems with \cite{paredes04} and without \cite{kinoshita04} 
an optical lattice along the 1D axes. 
On the other hand quantities like the NO and the OPDM still display the known 
behavior in equilibrium systems. In this way starting from a strongly 
interacting 1D Bose gas one can realize a very unconventional system of 
bosons displaying a Fermi edge on its momentum distribution function. 

We are grateful to P. Pedri for insightful discussions, and to M. A. 
Cazalilla for bringing to our attention Ref.\ \cite{sutherland98}. 
We thank HLR-Stuttgart for allocation of computer
time, and to SFB 382 for financial support.

\end{document}